\documentclass[12pt]{article}
\usepackage[a4paper]{geometry}
\usepackage{amsmath}
\usepackage{amssymb}
\usepackage[dvips]{graphicx}
\usepackage{psfig}

\begin{document}

\noindent
\begin{center}
{\large \bf The High Temperature Dynamics of a mean field Potts glass}\\[2mm]
Claudio Brangian$^*$, Walter Kob$^\ddagger$\footnote{Author to whom
correspondence should be addressed}, and Kurt Binder$^*$\\
$^*$ Institute of Physics, Johannes Gutenberg University, Staudinger Weg 7,\\
D-55099 Mainz, Germany\\
$\ddagger$ Laboratoire des Verres, Universit\'e Montpellier II, F-34095
Montpellier, France\\

\end{center}

\vspace*{7mm}
\par
\noindent
\begin{center}
\begin{minipage}[h]{122mm}
We use Monte Carlo simulations to investigate the dynamical properties
of the infinite range 10 states Potts glass. By analyzing the spin
autocorrelation function for system sizes up to $N=2560$, we show that
strong finite size effects are present around the predicted dynamical
transition temperature. The autocorrelation function shows strong
self-averaging at high temperatures, whereas close to the dynamical
transition they show the lack of self-averaging.

\end{minipage}
\end{center}

\vspace*{5mm}
\par
\noindent

\section{Introduction}
In recent years a new class of disordered spin glass models has been
introduced (for a review see Kirkpatrick and Thirumalai 1995) that show
strong analogies with the theoretical scenario proposed for the structural
glass transition, such as the presence of a dynamical transition at
a temperature $T_D$ and a static phase transition (with discontinuous
order parameter but without latent heat) at $T_0 < T_D$. Furthermore, the
equations of motion for the spin autocorrelation functions are formally
analogous to the equations of motion of the density-autocorrelation
functions introduced by the mode coupling theory of the glass transition
(G\"otze 1989). One example of such spin models is the $p$ states mean
field Potts glass with $p>4$ (Kirkpatrick and Wolynes 1987, Kirkpatrick
and Thirumalai 1988). The goal of the present paper is to compare the
relaxation dynamics of such a system with a {\it finite} size with
the dynamics of the system in the thermodynamic limit. The latter has
previously been determined at the level of one step replica symmetry
breaking (De Santis {\it et al.} 1995). Furthermore we investigate to
what extend the time correlation functions are self-averaging or not.

\section{Model and Simulations}
In the Potts model each spin $\sigma_i$ is a discrete variable that
can take one of $p$ different values: $\sigma_i \in \{1,\ldots,p\}$. The
Hamiltonian is given by
\begin{equation}
H=-\sum_{i,j}J_{ij}(p\delta_{\sigma_i \sigma_j}-1) ,
\label{eq1}
\end{equation}
i.e. each spin interacts with all the others. The coupling constants $J_{ij}$
are taken from a Gaussian distribution
\begin{equation}
P(J_{ij})=\frac{1}{\sqrt{2\pi }(\Delta J)} \exp \left[- \frac{
(J_{ij}-J_0)^2}{2(\Delta J)^2}\right] .
\label{eq2}
\end{equation}
We consider the case $p=10$ since for this case the static transition
has a strong first order character (jump of the order parameter from zero
to $q_0=0.452$) (De Santis {\it et al.} 1995). Note that for $p=2$ we
recover the Sherrington-Kirkpatrick model.  It has been shown that in
order to prevent the system from ordering ferromagnetically, a negative
value of $J_0$ has to be chosen (Elderfield and Sherrington 1983,
Gross \textit{et al.} 1985, Cwilich and Kirkpatrick 1989) and therefore
we have chosen $J_0=(3-p)/(N-1)$. The variance is set to $\Delta J =
(N-1)^{-1/2}$ and in the following we will set the Boltzmann's constant
$k_b=1$. With these units a numerical solution of the replica equations
at the level of one step replica symmetry breaking predicts a dynamical
transition at $T_D=1.142$ and a static transition from a paramagnet to
a spin glass at $T_0=1.131$ (De Santis \textit{et al}. 1995).

We have simulated 5 different system sizes, $N=160,$ 320, 640, 1280
and 2560 spins, at various temperatures. In this paper we will focus on
temperatures between $T=1.8$ to $T_D=1.142$, i.e. the range above the
dynamical transition temperature in the thermodynamic limit. Due to the
random nature of the interactions, we have to average all observables
not only over the canonical distribution but also over the possible
realizations of the disorder given by Eq.~(\ref{eq2}). In the following
we will denote this latter average by $\left [ \cdot \right]$. For
this we used $500$ different samples for $160$ spins, $100$ for $320,
640 \;\textrm{and}\; 1280$ spins and between $20 \;\textrm{and}\;
50$ for $2560$ spins.  The dynamics is generated using the Metropolis
algorithm. Starting from a given spin configuration, a spin is picked at
random and assigned a new random orientation. If the energy difference
between these two states is negative the move is accepted. If it is
positive it is accepted only with probability $\exp\left (-\Delta E/T
\right)$.

\section{Results}
We present now our results regarding the analysis of the spin-spin 
autocorrelation function, defined as

\begin{equation}
\label{ssaf}
C(t)= \frac{p}{p-1} \frac{1}{N}\sum_{i}^{N} \left[ \left\langle 
      \left( {\delta_{\sigma_i(t') \sigma_i(t'+t)}-1/p} \right) \right\rangle .
  \right]
\end{equation}

The mean field theory predicts that in the thermodynamic limit the
dynamics of the system slows down upon approach of the (dynamical)
transition temperature $T_D$. The relaxation time $\tau(T)$ for the time
correlation function should show at $T_D$ a divergence of the form $\tau
\propto (T-T_D)^{-\Delta}$. At the same time the spin autocorrelation
function is predicted to show at intermediate times a plateau with a
height $q_{EA}=0.328$ before it decays on the time scale $\tau$ towards
zero. At $T=T_D$ the system becomes nonergodic in that the correlation
function does not show anymore the final decay, i.e. it stays at
the plateau even for infinite times. (For a review of this behavior
on a class of spin glass systems see Thirumalai and Kirkpatrick 1995.
For the works regarding the Potts glass, see  Kirkpatrick and Wolynes,
1987 and  Kirkpatrick and Thirumalai, 1988).

However, for a {\it finite} system we do no longer expect the sharp
ergodic to nonergodic transition at $T_D$, since for $N< \infty$ the
relaxation times have to remain finite for all $T>0$. It is therefore
of interest to see how the typical relaxation behavior of the system
changes if $N$ is increased. 

\begin{figure}[f]
\psfig{file=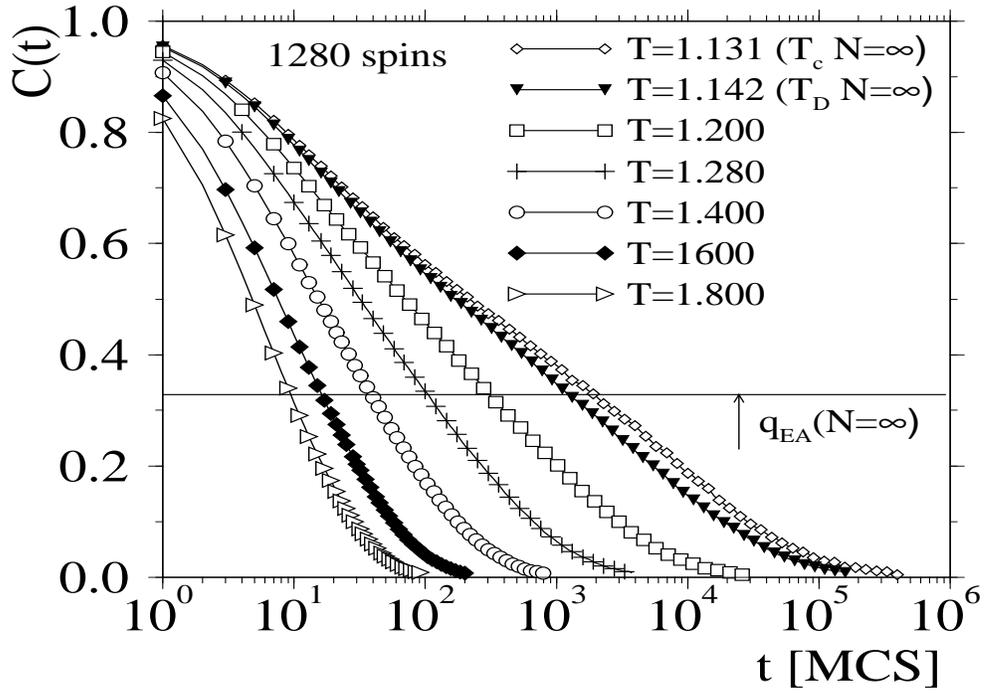,width=13cm,height=9.5cm}
\caption{Spin autocorrelation function $C(t)$ versus $t$ (measured 
in units of Monte Carlo steps per spin) for $1280$ spins at
various temperatures. The horizontal solid line shows the position of the
Edward Anderson order parameter in the thermodynamic limit.}
\label{fig1}
\end{figure}

In Fig. \ref{fig1} we show the temperature dependence of $C(t)$ for
$1280$ spins. From the figure we see that with decreasing temperature
the dynamics does indeed slow down. However, even at $T_0<T_D$ we do
not see a pronounced sign for the existence of the dynamic transition,
since the curves show, instead of the expected plateau at $q_{EA}$ (shown
in the figure as solid horizontal line), only a weak shoulder. Thus we
conclude that the dynamics of this system is strongly affected by finite
size effects even for systems as large as 1280 spins. The reason for this
is that the barriers in the free energy that separate one ``basin''
of configuration space from a neighboring ``basin'' are apparently
not very high, in contrast to the thermodynamics limit in which their
height diverges. Note that this $N$ dependence of the dynamics is in
marked contrast with the one found in structural glasses since in these
systems finite size effects are usually absent if the system has more
than a few hundred particles (Kob 1999).

\begin{figure}[f]
\psfig{file=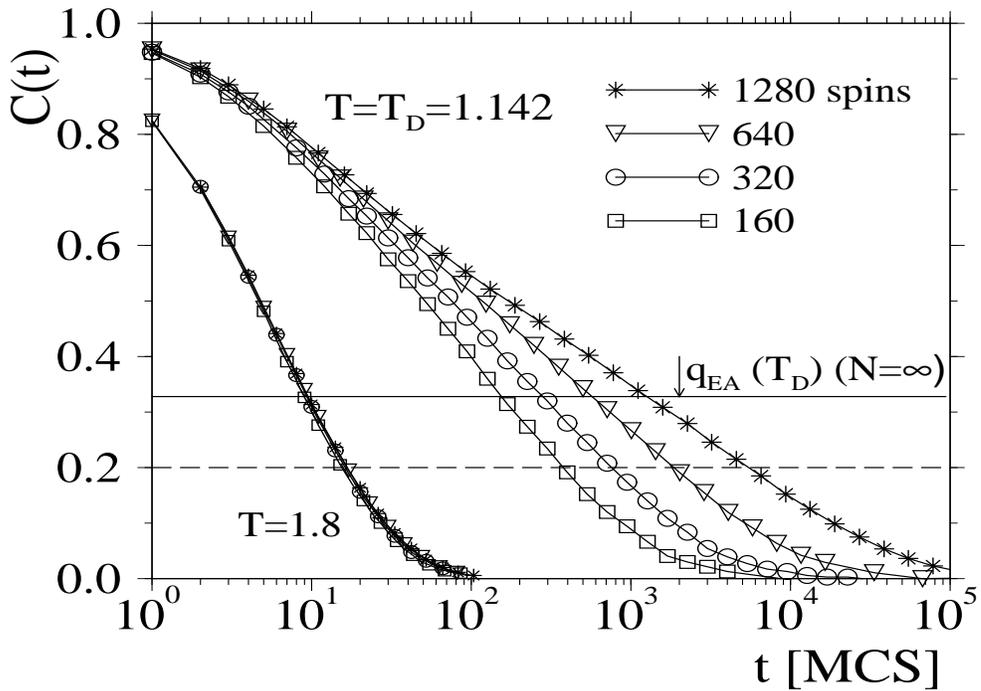,width=13cm,height=9.5cm}
\caption{Spin autocorrelation functions for different system sizes at two
different temperatures, $T=1.8$ and $T=1.142=T_D$. The horizontal solid line shows
the position of the Edward Anderson order parameter in the thermodynamic
limit. The dashed line is used to define the relaxation time $\tau(T)$. }
\label{fig2}
\end{figure}

To study in more detail how the relaxation dynamics depends on the size
of the system we show in Fig.~\ref{fig2} the time correlation function
for different system sizes for a high temperature, and at $T=T_D$. At the
high temperature the different $C(t)$ fall nicely onto a master curve,
i.e. there are no finite size effects. This is in stark contrast to
the behavior at $T_D$. At this temperature we find that the relaxation
dynamics for the different system sizes depends strongly on $N$ since
the large systems relax much slower than the small ones. If one defines a
relaxation time $\tau(T)$ via $C(\tau)=0.2$, see horizontal dashed line
in the figure, one finds that, at $T_D$, these times show a power-law
dependence: $\tau\propto N^{1.5}$ (Brangian \textit{et al.} 2001). Note
that the value of 1.5 for the exponent is significantly larger than the
estimate 2/3 for the Sherrington-Kirkpatrick model (Bhatt and Young 1992).

Using a dynamical finite size scaling Ansatz we have found that {\it
in the thermodynamic limit} we expect a divergence of the relaxation
time of the form $\tau \propto (T-T_D)^{-2}$ (Brangian \textit{et al.}
2001). Thus the exponent $-2$ is close to the values found for structural
glasses (G\"otze 1999).

In the remaining of this paper we will concentrate on the self-averaging
properties of $C(t)$. Knowing these properties will help to decide whether
or not it is necessary to average the results of a simulation over many
independent realizations of the disorder even in the case that the size
of the system is very large. Suppose that we have determined the thermal
average $X_i$ of an observable $X$. (Here $i$ stands for the realization
of the disorder.) Following Wiseman and Domany (1998, and references
therein) we consider the quantity $R_X$ defined as follows:
\begin{equation}
R_X=\frac{[X_i^2]-[X_i]^2}{[X_i]^2}.
\label{eq3}
\end{equation}
Here $[\cdot ]$ stands again for the average over the disorder. Usually
one has the situation of ``strong self averaging'' which means that
$R_X \propto 1/N$ for $N\gg 1$. The case $R_X \propto 1/N^{\alpha}$,
with $\alpha <1$ is denoted by ``weak self averaging''. Finally the case
$R_X = const.$ is called ``non self-averaging''.

\begin{figure}[f]
\psfig{file=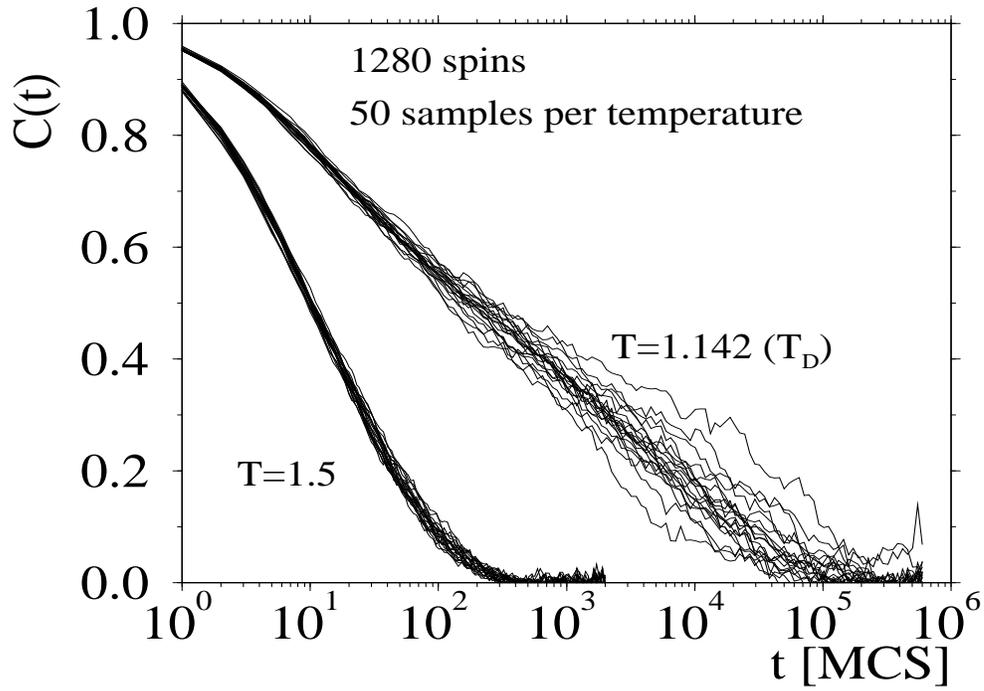,width=13cm,height=9.5cm}
\caption{Correlation functions for different realization of disorder.
         System with 1280 spins, temperature $T=1.5$ and $T=1.142=T_D$}
\label{fig3}
\end{figure}

In Fig.~\ref{fig3} we show the spin autocorrelation function for
system size $N=1280$ and for 20 representative samples. From the
figure it becomes clear that at high temperatures the sample to sample
fluctuations are quite small and that therefore the system is probably
self-averaging. For a temperature close to $T_D$ this is, however,
not the case in that the fluctuations are now on the same order as the
typical relaxation time.

\begin{figure}[f]
\psfig{file=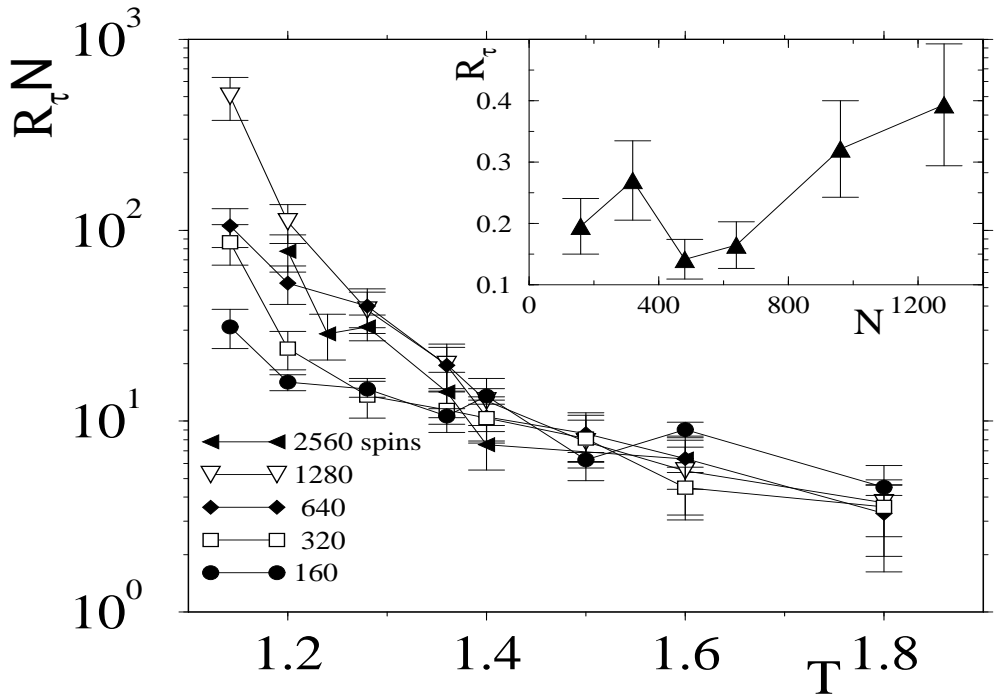,width=13cm,height=9.5cm}
\caption{Plot for the scaled quantity $R_X \cdot N$ as a function of 
         temperature; the inset shows $R_X$ as a function of the
         system size at $T=1.142=T_D$}
\label{fig4}
\end{figure}

To study this effect in a more quantitative way we use the
relaxation time $\tau$ as the observable $X$ discussed above. Using
thus equation~(\ref{eq3}) to define the quantity $R_{\tau}$ we can
investigate the $N$ dependence of $R_{\tau}$. In Fig.~\ref{fig4} we
show the temperature dependence of $R_{\tau}N$ for all system sizes
investigated (main figure), with error bars that have been determined with
the jackknife method (Newman and Barkema 1999).  We see that for high
temperatures we do indeed find that this quantity goes to a constant of
order one, {\it independent of the system size}.  Hence we conclude that
$R_{\tau}$ is proportional to $1/N$ and that hence the system is strong
self-averaging. At low temperatures this is, however, no longer the case
since there we see that the product increases with increasing system size
and becomes, for the largest systems, as large $O(10^3)$. Thus this is
evidence that the system is no longer self-averaging. To investigate this
point closer, we plot in the inset $R_{\tau}$ at $T_D$ as a function of
$N$. (Note that at this temperature we do not have data for the largest
system size since the relaxation time becomes too large.) From this graph
we see that the value of $R_{\tau}$ is basically constant within the
noise of the data, or shows even a slight trend to increase. Thus this
is evidence that at this temperature the system is not self-averaging. We
also mention that we expect that for sufficiently large $N$ self-averaging
will be recovered for all $T>T_D$, although our data are not conclusive
on this issue for $T\le 1.3$, due to the strong finite size effects.

To conclude, we have analyzed the dynamics of a $10$ states infinite
range Potts glass. Analytical results show that that this is a spin
model which resembles in many points structural glasses. We have shown
that the mean field scenario can, from a qualitative point of view, also
be seen in systems with finite $N$. However, close to the transition
temperature dynamical as well as static quantities are strongly
affected by finite size effects. In particular we find that the dynamics of the
system shows a crossover from a self-averaging behavior to a non self-averaging
behavior as the temperature approaches $T_D$.

{\bf Acknowledgements:}
C.B. was partially supported by the Deutsche Forschungsgemeinschaft,
Sonderforschungsbereich 262/D1. W.K. and K.B. are grateful to the
German Israeli foundation (GIF) for travel support. We thank the RUS
for a generous grant of computing time on the Cray T3E.

\newpage
\section*{References}
\begin{trivlist}
\item[]
Bhatt, R. N., and Young, A. P., 1992,
\textit{Europhys. Lett.}, \textbf{20}, 59.

\item[]
Brangian, C., Kob, W., and Binder, K., 2001,
\textit{Europhys. Lett.}, \textbf{56}, 756.

\item[]
Cwilich, G., and Kirkpatrick, T.R., 
1989, \textit{J. Phys. A}, \textbf{22}, 4971

\item[]
Elderfield, D., and Sherrington, D., 
1983, \textit{J. Phys. C}, \textbf{16}, L497

\item[]
De Santis, E., Parisi, G., and Ritort, F., 1995 
{\it J. Phys. A: Math. Gen}. {\bf 28} 3025

\item[]
G\"otze, W., 1989,
\textit{Liquids, freezing and the glass transition},
edited by J. P. Hansen, D. Levesque and J. Zinn-Justin,
(Amsterdam, North-Holland), pp. 287-503

\item[]
Gross, D.J., Kanter, I., and Sompolinsky, H.,
1985, \textit{Phys. Rev. Lett.}, \textbf{55}, 304

\item[]
Kirkpatrick, T. R., and Wolynes, P. G., 1987,
\textit{Phys. Rev. B}, \textbf{36}, 5388.

\item[]
Kirkpatrick, T. R., and Thirumalai, D., 1988,
\textit{Phys. Rev. B}, \textbf{37}, 5342.

\item[]
Kirkpatrick, T. R., and Thirumalai, D., 1995,
\textit{Transp. Theory Stat. Phys.}, \textbf{24}, 927.

\item[]
Kob, W, 1999,
\textit{J. Phys.: Condens. Matter}, \textbf{11}, R85.

\item[]
Newman, M. E. J., and Barkema, G. T., 1999,
\textit{Monte Carlo Methods in Statistical Physics}, 
(Oxford: Oxford University Press).

\item[]
Wiseman, S., and Domany, E., 1998,
\textit{Phys. Rev. E}, \textbf{58}, 2938.

\end{trivlist}

\newpage

\end{document}